\documentstyle[12pt,amssym]{article}

\def\C{\mbox{$\Bbb C$}}
\def\case#1#2{{\textstyle{#1\over #2}}}
\def\Ab{\bar{A}}
\def\Db{\bar{D}}

\title{
CREATION AND ANNIHILATION OPERATORS AND COHERENT STATES FOR THE
PT-SYMMETRIC OSCILLATOR}
\author{B. BAGCHI\\
{\small \sl Department of Applied Mathematics, University of Calcutta,} \\
{\small \sl 92 Acharya Prafulla Chandra Road, Calcutta 700 009, India}\\
{\small \sl E-mail: bbagchi@cucc.ernet.in}\\ [10pt] 
C. QUESNE\thanks{Directeur de recherches FNRS} \\
{\small \sl Physique Nucl\'eaire Th\'eorique et Physique Math\'ematique,}
\\ {\small \sl Universit\'e Libre de Bruxelles, Campus de la Plaine CP229,} \\ 
{\small \sl Boulevard~du Triomphe, B-1050 Brussels, Belgium} \\
{\small \sl E-mail: cquesne@ulb.ac.be}}
\date{ }
\begin{document}
\baselineskip=22pt plus 1pt minus 1pt
%%%%%%%%%%%%%%%%%%%%%%%%%%%%%%%%%%%%%%%%%%%%%%%%%%%%%%%%%%
\maketitle

\begin{abstract}
We construct two commuting sets of creation and annihilation operators for the
PT-symmetric oscillator. We then build coherent states of the latter as eigenstates of such
annihilation operators by employing a modified version of the normalization integral that
is relevant to PT-symmetric systems. We show that the coherent states are normalizable
only in the range (0, 1) of the underlying coupling parameter $\alpha$.
\end{abstract}

\vspace{1cm}
\noindent
Running head: PT-Symmetric Oscillator


\newpage
The current trend of interest in PT-symmetric non-Hermitian systems was triggered off
by an observation of Bender and Boettcher~\cite{bender} (see also Bessis~\cite{bessis})
that PT-invariance, in a number of cases, leads to energy eigenvalues that are real.
Recently, we studied~\cite{bagchi01a} a modified normalization condition for such
systems within the framework of a generalized continuity equation as a natural
replacement of the one prevailing for their Hermitian counterparts. In particular, for the
PT-symmetric oscillator (PTO)~\cite{znojil99}, we found that the modified normalization
integral of eigenfunctions ceases to be always positive definite. The purpose of this Letter
is twofold. First, we build two commuting sets of creation and annihilation operators for the
PTO problem, associated with the two quasi-parities $q = -1$ and $q = +1$, respectively.
Second, we construct normalizable coherent states (CS) for the same by taking into
account the corresponding pseudo-norm. Our analysis shows that the eigenstates $|z,
\alpha, q\rangle$ of the PTO annihilation operators, where the coupling parameter
$\alpha$ is related to the strength of the centrifugal-like term, have a form
similar to that of the standard CS of the Hermitian oscillator except for
the normalization coefficient. Actually, for $q= -1$ or $q = +1$ and $\alpha \in (0,
1)$, the normalization coefficient may be taken equal to that of the standard CS by
choosing the sign of the CS squared pseudo-norm appropriately. However, for $q = +1$
and $\alpha >1$, the dependence of the sign of the squared pseudo-norm of a Hamiltonian
eigenstate upon the choice of the latter has the effect that the CS squared pseudo-norm
vanishes for some value of $|z|$. Thus the corresponding CS are not normalizable. As a
result, normalizable CS exist only for $\alpha \in (0, 1)$.\par
%
%----------------------------------------------------------------------------------------------------------
%
We begin by writing down the Hamiltonian for the PTO that reads~\cite{znojil99}
\begin{equation}
  H^{(\alpha)} = - \frac{d^2}{dx^2} + (x - {\rm i}c)^2 + \frac{\alpha^2 - \frac{1}{4}}
  {(x - {\rm i}c)^2} \qquad \alpha > 0  \label{eq:H}
\end{equation}
for $\hbar = 2m = 1$. The form~(\ref{eq:H}), under the presence of a centrifugal-like
core of strength $G = \alpha^2 - \frac{1}{4}$, has a correspondence with the
Hamiltonian of the usual three-dimensional radial oscillator for a complex shift of
coordinate $r \to x - {\rm i}c$, $c>0$. However, the shift of the singularity off the path
of integration makes (\ref{eq:H}) exactly solvable on the entire real line for any $\alpha
> 0$. Note that $H^{(\alpha)}$ reduces to the Hamiltonian of the linear harmonic
oscillator for $\alpha = \frac{1}{2}$ and $c=0$.\par
%
%---------------------------------------------------------------------------------------------------------
% 
An important feature of the PT-symmetric Hamiltonian $H^{(\alpha)}$ is that, for
$\alpha$ different from an integer (which is assumed hereafter), it depicts a double
series of real energy eigenvalues $E^{(\alpha)}_{qn}$ distinguished by the quasi-parity
$q = \pm 1$. These energy levels read
\begin{equation}
  E^{(\alpha)}_{qn} = 4n + 2 - 2q\alpha \qquad n = 0, 1, 2, \ldots
\end{equation}
and point to a non-equidistant spectrum. The accompanying eigenfunctions are given
in terms of generalized Laguerre polynomials
\begin{equation}
  u^{(\alpha)}_{qn}(x) = {\cal N}^{(\alpha)}_{qn} e^{-\frac{1}{2}(x - {\rm i} c)^2} 
  (x - {\rm i} c)^{-q\alpha + \frac{1}{2}} L_n^{(-q \alpha)}[(x - {\rm i} c)^2]
  \label{eq:u}
\end{equation}
where ${\cal N}^{(\alpha)}_{qn}$ are the associated normalization coefficients. The
Laguerre polynomials are, of course, defined only for $-q\alpha > -1$, which, in turn,
implies the interval $0 < \alpha < 1$; however, their definition can be extended to any
value of $\alpha$.\par
%
%----------------------------------------------------------------------------------------------------------
%
In Ref.~\cite{bagchi01a}, we observed that PT-symmetric systems admit of a
generalized continuity equation that leads to a new definition of the normalization
condition of the bound-state eigenfunctions. The ingredients of the calculations are not
repeated here. We merely note that the pseudo-norm condition
\begin{equation}
  (u|u) \equiv \int_{-\infty}^{+\infty} dx\, u^*(-x) u(x) = \pm 1  \label{eq:ps-norm}
\end{equation}
serves as an appropriate counterpart of the standard normalization integral~\cite{schiff}
in PT-symmetric quantum mechanics. An offshoot of~(\ref{eq:ps-norm}) is that the
space of states on which PT-symmetric quantum mechanics manifests itself is not
necessarily the Hilbert space of quantum mechanics~\cite{japaridze, ahmed,
mostafazadeh}. The situation is akin to that in ordinary quantum mechanics where vectors
with negative norms are generally associated with an indefinite metric space~\cite{gupta}.
Note that the corresponding indefinite scalar product in PT-systems obeys $(u|v) =
(v|u)^*$.\par
%
%-------------------------------------------------------------------------------------------------------
%
When applied to the eigenfunctions~(\ref{eq:u}), the left-hand side
of~(\ref{eq:ps-norm}) can be evaluated to be
\begin{equation}
  I^{(\alpha)}_{qn} \equiv \bigl(u^{(\alpha)}_{qn} \big| u^{(\alpha)}_{qn}\bigr) =  
  \int_{-\infty}^{+\infty} dx\, \left[u^{(\alpha)}_{qn}(-x)\right]^* u^{(\alpha)}_{qn}(x)
  = \left|{\cal N}^{(\alpha)}_{qn}\right|^2 \cos \pi(-q\alpha + \case{1}{2})
  \frac{\Gamma(- q\alpha + n +1)}{n!}  \label{eq:PTO-norm}
\end{equation}
for $\alpha \in (0, 1)$. The presence of the cosine factor in the right-hand side
of~(\ref{eq:PTO-norm}) is a new feature, which signifies that for $\alpha \in (0,1)$, the
latter is positive ot negative according to whether $q = +1$ or $-1$, respectively.
Consistency with the condition~(\ref{eq:ps-norm}) then leads to the fixation
\begin{equation}
  \left|{\cal N}^{(\alpha)}_{qn}\right| = \left(\frac{n!}{\Gamma(- q\alpha + n +1) \cos \pi
  ( -\alpha + \case{1}{2})}\right)^{1/2}. 
\end{equation}
\par
%
%----------------------------------------------------------------------------------------------------------
% 
We now turn to the construction of CS associated with the PTO. To obtain CS, one should
consider the eigenstates of a ladder operator. The techniques of supersymmetric
quantum mechanics prove rather useful in that we are able to obtain commuting sets of
creation and annihilation operators~\cite{fukui}. To this end, we observe that there are
two independent complex superpotentials $W^{(\alpha)}_+(x)$ and
$W^{(\alpha)}_-(x)$ that accompany the Hamiltonian
$H^{(\alpha)}$~\cite{bagchi01b}. These are given by
\begin{equation}
  W^{(\alpha)}_q(x) = x - {\rm i}c + \frac{q\alpha - \frac{1}{2}}{x - {\rm i}c} \qquad q =
  \pm 1
\end{equation}
and satisfy the so-called shape-invariance condition
\begin{equation}
  \left(W^{(\alpha)}_q(x)\right)^2 + \frac{d}{dx} W^{(\alpha)}_q(x) =
  \left(W^{(\alpha-q)}_q(x)\right)^2 - \frac{d}{dx} W^{(\alpha-q)}_q(x) + 
  R^{(\alpha-q)}_q  \label{eq:shape}
\end{equation}
where $R^{(\alpha-q)}_q = 4$, $q = \pm 1$.\par
%
%--------------------------------------------------------------------------------------------------------
% 
Let $T$ be the operator that reparametrizes $\alpha$ by changing it into $\alpha-1$:
\begin{eqnarray}
  T |\phi^{(\alpha)}\rangle & = & |\phi^{(\alpha-1)}\rangle \\
  T O^{(\alpha)} T^{-1} & = & O^{(\alpha-1)} 
\end{eqnarray}
when acting on any suitable $\alpha$-dependent vector $|\phi^{(\alpha)}\rangle$ or
operator $O^{(\alpha)}$. Then the inverse operator $T^{-1}$ changes $\alpha$ into
$\alpha+1$.\par
%
%---------------------------------------------------------------------------------------------------------
%
Introducing now the operators $D^{(\alpha)}_q$ and $\Db^{(\alpha)}_q$ defined by
\begin{equation}
  D^{(\alpha)}_q = \frac{d}{dx} + W^{(\alpha)}_q(x) \qquad
  \Db^{(\alpha)}_q = - \frac{d}{dx} + W^{(\alpha)}_q(x) 
\end{equation}
the shape-invariance condition~(\ref{eq:shape}) can be reset as
\begin{equation}
  D^{(\alpha)}_q \Db^{(\alpha)}_q = T^q \left(\Db^{(\alpha)}_q D^{(\alpha)}_q +
  R^{(\alpha)}_q\right) T^{-q}. 
\end{equation}
This form suggests that if we define two new operators
\begin{equation}
  A^{(\alpha)}_q = \case{1}{2} T^{-q} D^{(\alpha)}_q \qquad \Ab^{(\alpha)}_q =
  \case{1}{2} \Db^{(\alpha)}_q T^q
\end{equation}
the shape-invariance condition is automatically transformed into a Heisenberg-Weyl-like
algebra
\begin{equation}
  \left[A^{(\alpha)}_q, \Ab^{(\alpha)}_q\right] = \case{1}{4} R^{(\alpha)}_q = 1
  \qquad q = \pm 1.  \label{eq:HW}
\end{equation}
Up to a lack of Hermiticity, equation~(\ref{eq:HW}) furnishes two sets of creation and 
annihilation operators for the PTO, which, to the best of our knowledge, are being
constructed for the first time here. In this connection it may be pointed out that those
obtained in~\cite{znojil00} are actually raising and lowering generators of two sl(2)
algebras. In the present case, it is quite straightforward to establish that the operators
corresponding to both quasi-parities commute with one another:
\begin{equation}
  \left[A^{(\alpha)}_+, A^{(\alpha)}_-\right] = \left[A^{(\alpha)}_+,
  \Ab^{(\alpha)}_-\right] = \left[\Ab^{(\alpha)}_+, A^{(\alpha)}_-\right] =
  \left[\Ab^{(\alpha)}_+, \Ab^{(\alpha)}_-\right] = 0.  
\end{equation}
\par
%
%-----------------------------------------------------------------------------------------------------------
%
It may be remarked that for each $q$ value the lowest eigenstate
$\left|u^{(\alpha)}_{q0}\right\rangle$ of $H^{(\alpha)}$ is annihilated by the
corresponding $D^{(\alpha)}_q$:
\begin{equation}
  A^{(\alpha)}_q \left|u^{(\alpha)}_{q0}\right\rangle = 0.
\end{equation}
Further, for each $q$ value we can construct a sequence of Hamiltonians
\begin{eqnarray}
  H^{(\alpha)}_{qn} & \equiv & 4 A^{(\alpha - nq)}_q \Ab^{(\alpha - nq)}_q + 4(n-1)
        \nonumber \\
  & = & 4 \Ab^{(\alpha - nq)}_q A^{(\alpha - nq)}_q + 4n \nonumber \\
  & = & H^{(\alpha - nq)} + 2q\alpha + 2n - 2 \qquad n=0, 1, 2, \ldots
\end{eqnarray}
as long as $\alpha - nq$ remains positive. The Hamiltonians $H^{(\alpha)}_{qn}$ and
$H^{(\alpha)}_{q,n+1}$ are superpartners in the sense that they are isospectral except
for the lowest eigenstate of $H^{(\alpha)}_{qn}$ with quasi-parity~$q$.\par
%
%--------------------------------------------------------------------------------------------------------
% 
The eigenvalues of $H^{(\alpha)}_{qn}$ of quasi-parity $q$ are given by
\begin{equation}
  {\cal E}^{(\alpha)}_{qn} = \sum_{k=1}^n R^{(\alpha - kq)}_q = 4n
\end{equation}
and correspond to $E^{(\alpha)}_{qn} = {\cal E}^{(\alpha)}_{qn} - 2q\alpha + 2$. The
corresponding eigenvectors $\left|u^{(\alpha)}_{qn}\right\rangle$ are proportional to
$\Db^{(\alpha)}_q \Db^{(\alpha-q)}_q \ldots \Db^{(\alpha - nq + q)}_q
\left|u^{(\alpha - nq)}_{q0}\right\rangle$.\par
%
%----------------------------------------------------------------------------------------------------------
%  
Returning now to the PTO normalization integral~(\ref{eq:PTO-norm}), we note that, by
analytic continuation, it also holds for any complex value of $\alpha$. For the real
positive values of $\alpha$ considered here, this analytic continuation has the following
implications. For $q=-1$ and $N < \alpha < N+1$ ($N=0$, 1, 2,~\ldots),
$I^{(\alpha)}_{-n}$ has the sign $(-1)^{N+1}$ independently of $n$, while for $q=+1$
and $N < \alpha < N+1$, $I^{(\alpha)}_{+n}$ has the sign $(-1)^n$ if $n \le N-1$ and
$(-1)^N$ if $n \ge N$. We therefore normalize the eigenfunctions
$u^{(\alpha)}_{qn}(x)$ according to the prescription ($N < \alpha < N+1$, $N \ge 1$)
\begin{equation}
  \left( u^{(\alpha)}_{qn} \Big| u^{(\alpha)}_{qn}\right) = \sigma^{(\alpha)}_{qn}
  \label{eq:PTO-norm-bis} 
\end{equation}
where
\begin{equation}
  \sigma^{(\alpha)}_{qn} = \left\{\begin{array}{ll}
        (-1)^{N+1} & {\rm if\ } q=-1 \\[0.2cm]
        (-1)^{n} & {\rm if\ } q=+1 {\rm\ and\ } n \le N-1 \\[0.2cm]     
        (-1)^{N} & {\rm if\ } q=+1 {\rm\ and\ } n \ge N
  \end{array}\right..
\end{equation}
In the context of (\ref{eq:PTO-norm-bis}) the normalization coefficients ${\cal
N}^{(\alpha)}_{qn}$ transform as
\begin{equation}
  \left| {\cal N}^{(\alpha)}_{qn}\right| = \left\{\begin{array}{ll}
        \left[(-1)^N \sin\pi\alpha \frac{\Gamma(-q\alpha+n+1)}{n!}\right]^{-1/2} & {\rm
             if\ } q=-1 {\rm\ or\ } q=+1 {\rm\ and\ } n \ge N \\[0.2cm]
        \left[(-1)^n \sin\pi\alpha \frac{\Gamma(-\alpha+n+1)}{n!}\right]^{-1/2} & {\rm
             if\ } q=+1 {\rm\ and\ } n \le N-1 
  \end{array}\right.. 
\end{equation}
\par
%
%---------------------------------------------------------------------------------------------------------
%  
The actions of $A^{(\alpha)}_q$ and $\Ab^{(\alpha)}_q$ on the corresponding
normalized eigenvectors are given by
\begin{eqnarray}
  \Ab^{(\alpha)}_q \left|u^{(\alpha)}_{qn}\right\rangle  & = & \case{1}{2}
       \Db^{(\alpha)}_q T^q \left|u^{(\alpha)}_{qn}\right\rangle = \case{1}{2}
       \Db^{(\alpha)}_q \left|u^{(\alpha-q)}_{qn}\right\rangle = C^{(\alpha)}_{q,n+1}   
       \left|u^{(\alpha)}_{q,n+1}\right\rangle \\
  A^{(\alpha)}_q \left|u^{(\alpha)}_{q,n+1}\right\rangle  & = & \case{1}{2} T^{-q}
       D^{(\alpha)}_q \left|u^{(\alpha)}_{q,n+1}\right\rangle = T^{-q}
       C^{\prime(\alpha-q)}_{q,n+1} \left|u^{(\alpha-q)}_{qn}\right\rangle =
       C^{\prime(\alpha)}_{q,n+1} \left|u^{(\alpha)}_{qn}\right\rangle 
\end{eqnarray}
where the constants $C^{(\alpha)}_{q,n+1}$ and $C^{\prime(\alpha)}_{q,n+1}$ turn
out to be equal
\begin{equation}
  C^{(\alpha)}_{q,n+1} = C^{\prime(\alpha)}_{q,n+1} = \sqrt{n+1} \label{eq:C}
\end{equation}
and indeed independent of either $\alpha$ or $q$. To arrive at the results~(\ref{eq:C}),
we consider the scalar product $\left(\Db^{(\alpha)}_q u^{(\alpha-q)}_{qn} \Big|
\Db^{(\alpha)}_q u^{(\alpha-q)}_{qn}\right)$ and use the fact that
$\Db^{(\alpha)}_q$ is PT-antisymmetric. Integration by parts then leads to
\begin{equation}
  \left| C^{(\alpha)}_{q,n+1}\right|^2 \sigma^{(\alpha)}_{q,n+1} = - (n+1)
  \sigma^{(\alpha-q)}_{qn} 
\end{equation}
which gives $C^{(\alpha)}_{q,n+1}$ from the consideration that 
$\sigma^{(\alpha)}_{q,n+1} = - \sigma^{(\alpha-q)}_{qn}$.\par
%
%-------------------------------------------------------------------------------------------------------
%
On the other hand, the constant $C^{\prime(\alpha)}_{q,n+1}$ is obtained from
$C^{(\alpha)}_{q,n+1}$ by observing
\begin{equation}
  C^{\prime(\alpha-q)}_{q,n+1} \sigma^{(\alpha-q)}_{qn} = \case{1}{2} 
  \left(u^{(\alpha-q)}_{qn} \Big| D^{(\alpha)}_q u^{(\alpha)}_{q,n+1}\right) =
  - \case{1}{2} \left(\Db^{(\alpha)}_q u^{(\alpha-q)}_{qn} \Big| 
  u^{(\alpha)}_{q,n+1}\right) = - \left(C^{(\alpha)}_{q,n+1}\right)^*
  \sigma^{(\alpha)}_{q,n+1}  
\end{equation}
so that 
\begin{equation}
  C^{\prime(\alpha-q)}_{q,n+1} = \left(C^{(\alpha)}_{q,n+1}\right)^*.  \label{eq:C-C'}
\end{equation}
It is not difficult to see that we can always choose the phase of the eigenfunctions, that is
of ${\cal N}^{(\alpha)}_{qn}$, in such a way that (\ref{eq:C-C'}) is real and positive. The
role of the operators $\Ab^{(\alpha)}_q$ and $A^{(\alpha)}_q$ can thus be
summarized as
\begin{equation}
  \Ab^{(\alpha)}_q \left|u^{(\alpha)}_{qn}\right\rangle = \sqrt{n+1}\, 
  \left|u^{(\alpha)}_{q,n+1}\right\rangle \qquad A^{(\alpha)}_q
  \left|u^{(\alpha)}_{qn}\right\rangle = \sqrt{n}\, 
  \left|u^{(\alpha)}_{q,n-1}\right\rangle  \label{eq:ladder} 
\end{equation}
where $\left|u^{(\alpha)}_{qn}\right\rangle = \frac{1}{\sqrt{n!}}
\left(\Ab^{(\alpha)}_q\right)^n \left|u^{(\alpha)}_{q0}\right\rangle$.\par
%
%----------------------------------------------------------------------------------------------------------
%
With the ladder operator relations~(\ref{eq:ladder}) at hand, we now look for the
eigenstates $|z, \alpha, q\rangle$ of the annihilation operator $A^{(\alpha)}_q$
corresponding to the eigenvalues $z \in\C$:
\begin{equation}
  A^{(\alpha)}_q |z, \alpha, q\rangle = z |z, \alpha, q\rangle \qquad q = ±\pm 1.  
\end{equation}
Because of~(\ref{eq:ladder}), it is easy to construct the CS in terms of the Hamiltonian
eigenstates. They have a form similar to that of the standard CS of the Hermitian
oscillator except for the normalization coefficient:
\begin{equation}
  |z, \alpha, q\rangle = N^{(\alpha)}_q(|z|) \exp\left(z \Ab^{(\alpha)}_q\right)
  \left|u^{(\alpha)}_{q0}\right\rangle = N^{(\alpha)}_q(|z|) \sum_{n=0}^{\infty}
  \frac{z^n}{\sqrt{n!}} \left|u^{(\alpha)}_{qn}\right\rangle. 
\end{equation}
\par
%
%---------------------------------------------------------------------------------------------------------
%
The overlap of two such CS is given by
\begin{equation}
  (z', \alpha, q'|z, \alpha, q) = \delta_{q',q} \left[N^{(\alpha)}_q(|z'|)\right]^*
  N^{(\alpha)}_q(|z|) \sum_{n=0}^{\infty} \frac{(z^{\prime*}z)^n}{n!} 
  \sigma^{(\alpha)}_{qn} 
\end{equation}
where
\begin{eqnarray}
  \sum_{n=0}^{\infty} \frac{(z^{\prime*}z)^n}{n!} \sigma^{(\alpha)}_{-n} & = &
         (-1)^{N+1} \exp(z^{\prime*}z) \\
  \sum_{n=0}^{\infty} \frac{(z^{\prime*}z)^n}{n!} \sigma^{(\alpha)}_{+n} & = &
         \sum_{n=0}^{N-1} (-1)^n \frac{(z^{\prime*}z)^n}{n!} + (-1)^N 
         \sum_{n=N}^{\infty} \frac{(z^{\prime*}z)^n}{n!} \nonumber \\
  & = & (-1)^N \left\{\exp(z^{\prime*}z) + \sum_{n=0}^{N-1} \left[(-1)^{n-N} -1\right]
         \frac{(z^{\prime*}z)^n}{n!}\right\}  \label{eq:summation}   
\end{eqnarray}
and we have assumed $N < \alpha < N+1$. The summation portion in the right-hand side
of equation~(\ref{eq:summation}) contributes only when $n-N$ is odd.\par
%
%----------------------------------------------------------------------------------------------------------
%
{}For the normalization, we obtain
\begin{eqnarray}
  (z, \alpha, q|z, \alpha,q) & = & \left|N^{(\alpha)}_q(|z|)\right|^2 \Biggl\{\delta_{q,-1}
  (-1)^{N+1} \exp(|z|^2) \nonumber \\ 
  && \mbox{} + \delta_{q,+1} (-1)^N \Biggl[\exp(|z|^2) + \sum_{n=0}^{N-1} 
  \left((-1)^{n-N} -1\right)\frac{(z^{\prime*}z)^n}{n!}\Biggr]\Biggr\}. 
\end{eqnarray}
For $q=-1$, we may choose $(z, \alpha, -|z, \alpha,-) = (-1)^{N+1}$, thus giving
\begin{equation}
  \left|N^{(\alpha)}_-(|z|)\right| = \exp(-|z|^2/2)  \label{eq:CS-norm}
\end{equation}
as for the standard CS. Equation~(\ref{eq:CS-norm}) is valid independently of the value
of $N$, i.e., of $\alpha$. A result similar to~(\ref{eq:CS-norm}) can also be obtained for
$q=+1$ and $N=0$ in which case we may choose $(z, \alpha, +|z, \alpha, +) = 1$.\par
%
%---------------------------------------------------------------------------------------------------------
%
However, for $N\ge 1$, the CS with $q = +1$ are not normalizable for some value of
$|z|$. For instance, if $N=1$, we obtain
\begin{equation}
  (z, \alpha, +|z, \alpha, +) = - \left|N^{(\alpha)}_+(|z|)\right|^2 [\exp(|z|^2) - 2]
\end{equation}
where the second factor vanishes for $|z| = \sqrt{\ln2}$.\par
%
%--------------------------------------------------------------------------------------------------------
%
We are thus led to the conclusion that only for $0 < \alpha < 1$, the CS of the PTO
problem are normalizable. \par
%
%----------------------------------------------------------------------------------------------------------
%
The authors would like to thank M.\ Znojil for some interesting discussions on the
normalization problem for PT-symmetric systems.\par
%
%=============================================================
%

\end{document}